\documentclass[aps,pra,preprint,groupedaddress,showpacs,%
tightenlines,
nofootinbib,floatfix]{revtex4}

\usepackage{amsmath,amsfonts,amssymb,bm}
\usepackage{graphicx}
\usepackage{array,dcolumn}

\newcommand{\upd}{\mathrm{d}}
\newcommand{\chem}[1]{\ensuremath{\mathrm{#1}}}
\newcommand{\un}[1]{\ensuremath{\unskip\,\mathrm{#1}}}
\newcommand{\etal}{\unskip\ \emph{et al.}}

\begin{document}

\title{Diffusion radius of muonic hydrogen atoms in H-D gas}

\author{Andrzej Adamczak}
\email{andrzej.adamczak@ifj.edu.pl}
\affiliation{                    
  Institute of Nuclear Physics, Polish Academy of Sciences, 
  PL-31342~Krak\'ow, Poland}
\affiliation{
  Rzesz\'ow Technical University, PL-35959~Rzesz\'ow, Poland}

\author{Jakub Gronowski}
\email{jakub.gronowski@ifj.edu.pl}
\affiliation{                    
  Institute of Nuclear Physics, Polish Academy of Sciences, 
  PL-31342~Krak\'ow, Poland}

\date{\today}

\begin{abstract}
  The diffusion radius of the $1S$ muonic hydrogen atoms in gaseous
  \chem{H_2} targets with various deuterium admixtures has been
  determined for temperatures $T=30$ and~300\un{K}. The Monte Carlo
  calculations have been performed using the partial differential cross
  sections for $p\mu$ and $d\mu$ atom scattering from the molecules
  \chem{H_2}, \chem{HD} and~\chem{D_2}. These cross sections include
  hyperfine transitions in the muonic atoms, the muon exchange between
  the nuclei $p$ and~$d$, and rotational-vibrational transitions in the
  target molecules. The Monte Carlo results have been used for preparing
  the time-projection chamber for the high-precision measurement of the
  nuclear $\mu^{-}$~capture in the ground-state $p\mu$ atom, which is
  now underway at the Paul Scherrer Institute.
\end{abstract}


\maketitle


Theoretical studies of the muonic atom diffusion in molecular
hydrogen-isotope targets are important for many experiments in
low-energy muon physics. In particular, knowledge of the diffusion
radius of muonic hydrogen atoms in gaseous H-D targets is required for
investigations of the $\mu^{-}$ nuclear capture in the $p\mu$ and $d\mu$
atoms created in H-D targets. The diffusion radius~$R_\text{diff}$ is
defined as the distance between the point of the muon stop in~H-D and
the point of the muonic atom disappearance due to the muon decay or to
the muon nuclear capture. Since the $\mu^{-}$~capture rate on~$p$ or~$d$
is several orders of magnitude lower than the muon decay rate,
$R_\text{diff}$ is practically determined by the point of the muon
decay. A~high-precision measurement of the rate~$\Lambda_s$ for the muon
capture $p\mu\to{}\nu_\mu+n$ in the ground-state $p\mu$~atom (MuCap
experiment) is underway at the Paul Scherrer
Institute~\cite{kamm00,kamm01,kamm02,laus05}. The rate~$\Lambda_s$ for
the singlet state~$F=0$ of the total muonic atom spin~$F$ is sensitive
to the weak form factors of the nucleon, especially to the induced
pseudoscalar coupling constant~$g_P$. As a~result, this experiment will
provide a~rigorous test of theoretical predictions based on the Standard
Model and low-energy effective theories
of~QCD~\cite{gorr04,gova00}. A~high-precision measurement of the
$\mu^{-}$ capture rate in the process $d\mu\to{}\nu_\mu+n+n$ is under
consideration by the MuCap collaboration~\cite{kamm02}. Such an
experiment would be uniquely suited to study the axial meson exchange
currents in the two-nucleon system.

In this paper, main results of the Monte Carlo simulations for
determining the optimal conditions for the MuCap experiment are
presented. The time-projection chamber is filled with almost pure
\chem{H_2} gas which, however, contains a~very small \chem{D_2}
contamination. In the isotope exchange process $p\mu+d\to{}d\mu+p$, the
energy of about 135\un{eV} is released in the centre-of-mass system.
Therefore, the created $d\mu$ atom gains the collision energy of a~few
tens\un{eV}. As a~result, the diffusion radius is significantly
enlarged.  This leads to an enhanced absorption of the muons in the
time-projection-chamber walls and limits the spatial resolution. The
determination of the highest acceptable \chem{D_2} contamination has
been one of the aims of the presented simulations. Since the capture
rate~$\Lambda_s$ depends strongly on the total $p\mu$ spin, it is
necessary to calculate the time evolution of the population of the
$p\mu$ spin states. The initial distribution of the spin states $F=1$
and $F=0$ is statistical. The simulations have been performed for the
target temperatures $T=30$ and~300\un{K}. The target density has been
fixed at the constant value $\phi=0.01$ (relative to the liquid hydrogen
density of $4.25\times{}10^{22}\un{atoms/cm^3}$), which corresponds to
the pressure of about 9\un{bar} at~300\un{K}. At such a~density, the
probability of formation of the muonic molecule $pp\mu$ is small. In
higher-density targets, the muon nuclear capture inside~$pp\mu$ is
significant. This leads to serious problems with interpreting the
experimental data owing to inaccuracy of the rate for the ortho-para
conversion of the $pp\mu$
molecules~\cite{kamm00,kamm01,kamm02,laus05,gorr04}.  The spin-flip
transition $p\mu(F=1)+p\to{}p\mu(F=0)+p$ due to the muon exchange
between the protons is still sufficiently strong at $\phi\approx{}0.01$
to ensure a~fast quenching of the higher hyperfine state $F=1$ and,
therefore, an unambiguous $\Lambda_s$~measurement.

The Monte Carlo kinetics code includes the muon decay, $p\mu$ and $d\mu$
scattering from the molecules \chem{H_2}, \chem{HD} and~\chem{D_2}, and
formation of the molecules $pp\mu$, $pd\mu$ and $dd\mu$. In the
scattering process, the atoms can change their spin states. The isotope
exchange reaction $p\mu+d\to{}d\mu+p$ in $p\mu$ scattering from
\chem{HD} and~\chem{D_2} is taken into account. Also, all possible
rotational and vibrational transitions in the target molecules are
included. At the collision energies
$\varepsilon\lesssim\nobreak{}10\un{eV}$ (in the laboratory system), the
scattering processes are described using the differential cross sections
$\upd\sigma/\upd\Omega$ for scattering from the hydrogenic
molecules~\cite{adam93,adam06} (``molecular'' cross sections). At higher
energies, effects of the molecular binding and electron screening can be
neglected and, therefore, the differential cross sections for the muonic
atom scattering from hydrogen-isotope nuclei are
used~\cite{brac89,brac89a,brac90,chic92} (``nuclear'' cross sections).
\begin{figure}[htb]
  \includegraphics[height=5cm]{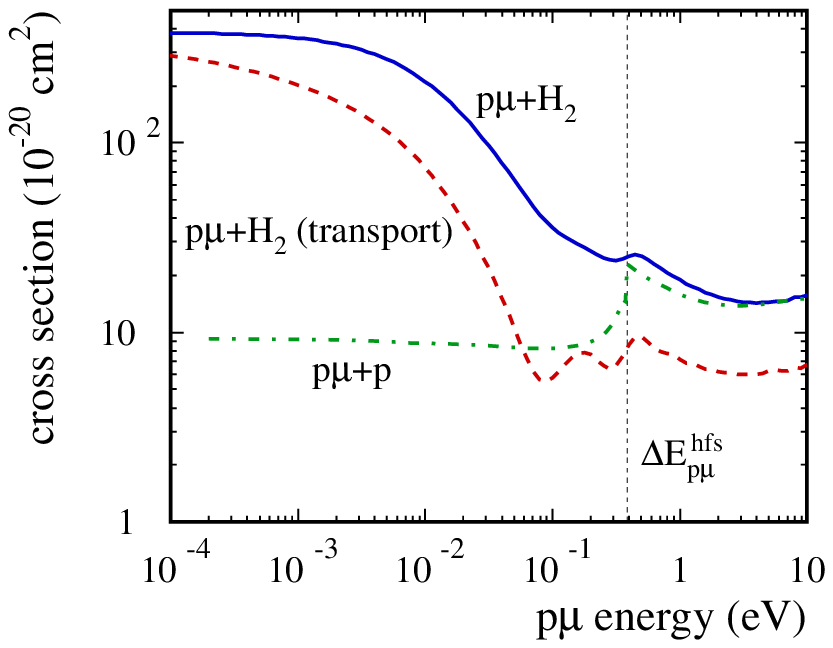}
  \includegraphics[height=5cm]{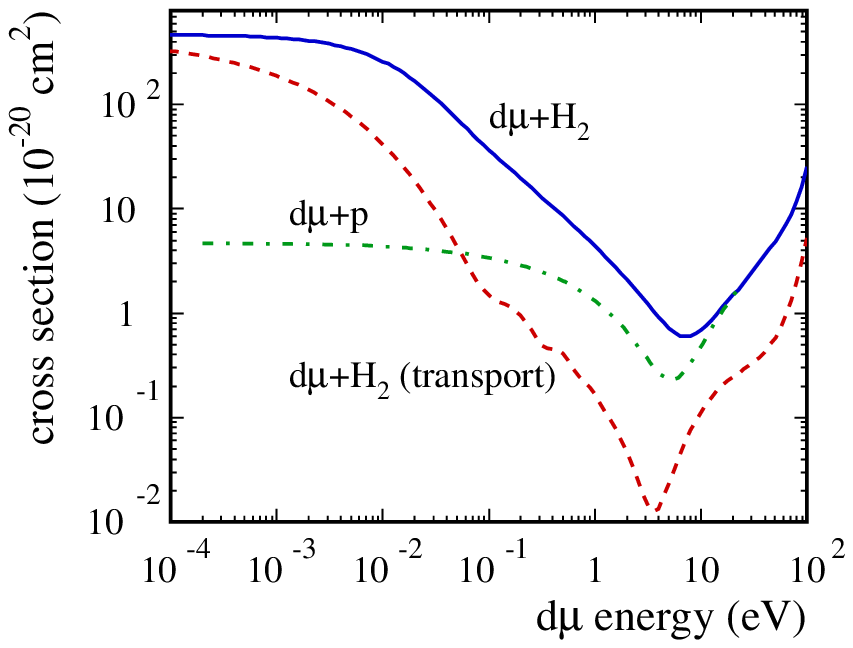}
  \caption{Transport (dashed lines) and total (solid lines) cross
    sections for the scattering of $p\mu(F=0)$ and $d\mu$ atoms from
    a~ground-state \chem{H_2}~molecule versus the collision
    energy~$\varepsilon$ in the laboratory system. The doubled total
    cross sections (dash-dotted lines) for the corresponding nuclear
    scattering are shown for comparison. The hyperfine-transition
    threshold is denoted by~$\Delta{}E^\text{hfs}_{p\mu}$.}
  \label{fig:xgtran}
\end{figure}
In fig.~\ref{fig:xgtran}, the total molecular cross sections for
$p\mu(F=0)$ and $d\mu$ scattering from the ground-state \chem{H_2}
molecule are shown as an example. The muonic atom spin is conserved in
the presented processes. Also, the corresponding transport cross
sections, defined as
\begin{equation}
  \label{eq:x_tran}
  \sigma_\text{tran} = \int \upd\Omega \, 
   (1 - \cos\vartheta) \frac{\upd\sigma(\vartheta)}{\upd\Omega} \,,
\end{equation}
are shown. The scattering angle is denoted here by~$\vartheta$. The
doubled total nuclear cross sections for the processes
$p\mu(F=0)+p\to{}p\mu(F=0)+p$ and $d\mu+p\to{}d\mu+p$ are plotted for
comparison. The transport cross sections approach the total cross
section only at $\varepsilon\to{}0$, which demonstrates strong
anisotropy of the molecular cross sections. Large differences between
the molecular and nuclear cross sections at
$\varepsilon\lesssim{}1\un{eV}$ are due to molecular-binding and
electron-screening effects. The total molecular and nuclear cross
sections for all combinations of the three hydrogen isotopes are
presented in~ref~\cite{adam96}.

The time evolution of the hyperfine states, the energy distribution of
the muonic atoms, and the radial distribution of the muon decays were
calculated for various initial conditions. All the presented results are
given for a~fixed target density $\varphi=0.01$. The initial
distribution of the $p\mu$ or $d\mu$ kinetic energy was described by the
two Maxwell components: thermal~(50\%) and energetic~(50\%) with the
mean energy~$\varepsilon_\text{avg}=$~1--5\un{eV}, according to the
experimental results~\cite{abbo97,wert96}.
\begin{figure}[htb] 
  \includegraphics[height=5cm]{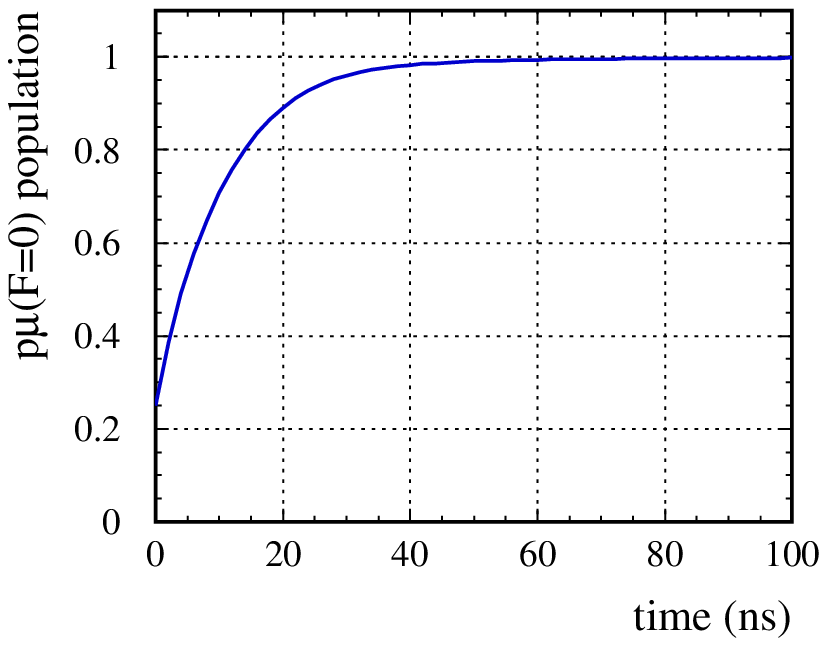}
  \includegraphics[height=5cm]{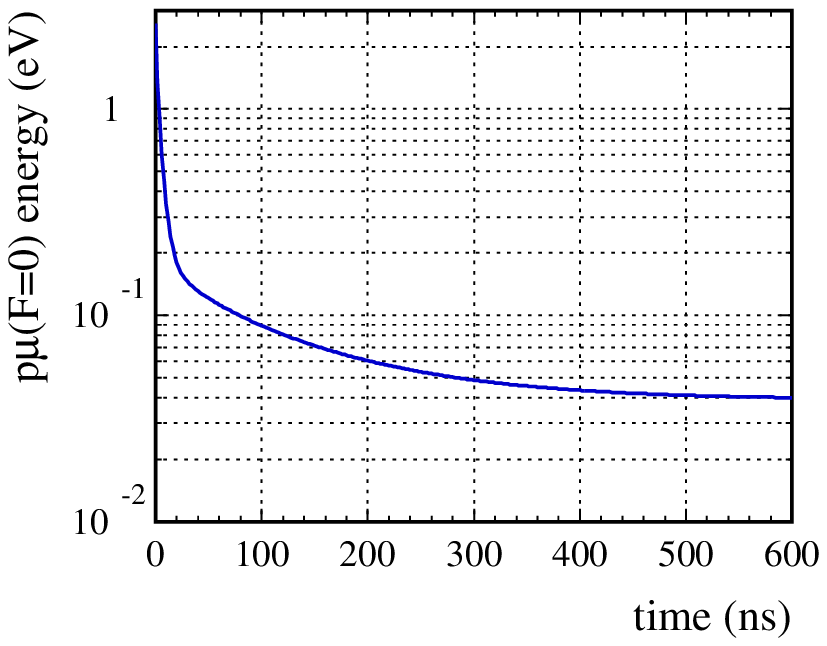}
  \caption{Time dependence of the $p\mu(F=0)$ population and of the mean
    $p\mu(F=0)$ kinetic energy~$\varepsilon_\text{avg}$ in a~pure
    \chem{H_2} at $T=300\un{K}$ and~$\varphi=0.01$.}
  \label{fig:enspi_pm300}
\end{figure}
The calculated time evolution of the $F=0$ state and of the mean
$p\mu(F=0)$ kinetic energy are shown in~fig.~\ref{fig:enspi_pm300}, for
a~pure~\chem{H_2} at $T=300\un{K}$. The $p\mu$ atoms starting at
$\varepsilon\sim{}1\un{eV}$ are slowed down within a~few tens\un{ns} to
energies where the spin-flip transitions $F=0\to{}F=1$ are impossible.
The hyperfine-transition threshold is
$\Delta{}E^\text{hfs}_{p\mu}=0.182\un{eV}$ in the $p\mu+p$
centre-of-mass system. After this time, the $F=1$ state disappears with
a~time constant of~6\un{ns}. Hence, about 50\un{ns} after the muon stop,
the relative population of the $F=1$ state is below~0.01 and the
measurement is no longer distorted by the population of the upper
hyperfine level. All that takes place when most of the initially
energetic atoms remains epithermal ($\varepsilon\gg{}k_\text{B}T$, where
$k_\text{B}$ is the Boltzmann constant). The $p\mu(F=0)$ thermalization
from $\varepsilon\approx{}0.1\un{eV}$ takes about~400\un{ns}.
\begin{figure} 
  \includegraphics[height=5cm]{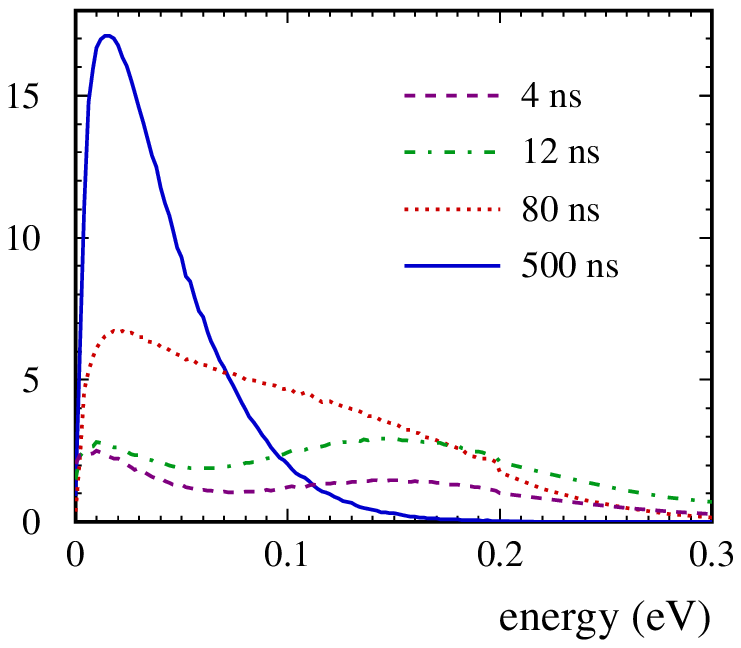}
  \includegraphics[height=5cm]{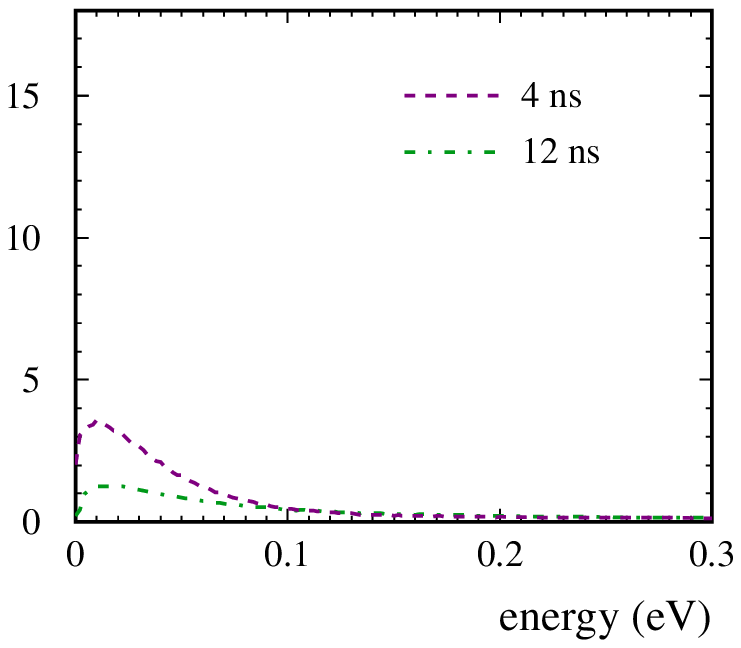}
  \caption{Energy distribution of $p\mu(F=0)$ and $p\mu(F=1)$ atoms in
    a~\chem{H_2} gas at $T=300\un{K}$, for several moments after the
    muon stop.}
  \label{fig:tedis_pm300}
\end{figure}
As it is illustrated in~fig.~\ref{fig:tedis_pm300}, the $p\mu(F=0)$
energy spectrum is epithermal for times much longer than in the case of
$p\mu(F=1)$ atoms. Only after the total deexcitation of the $F=1$ level,
the $p\mu(F=0)$ energy distribution takes the final Maxwellian form
with~$\varepsilon_\text{avg}=0.04\un{eV}$. Most of the $p\mu$ diffusion
until the muon decay takes place after the system has been thermalized.

\begin{figure} 
  \includegraphics[height=5cm]{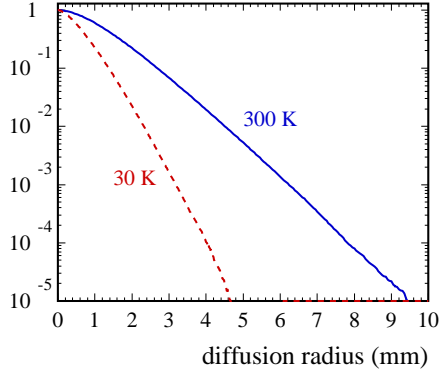}
  \caption{Fraction of the $\mu^{-}$~decays \textit{outside} the $p\mu$
    diffusion radius from the point of $p\mu$ formation in
    a~pure~\chem{H_2}, for times $t\leq{}20\un{\mu{}s}$ and $T=30$
    and~300\un{K}.}
  \label{fig:rdiff}
\end{figure} 
\begin{figure}
  \includegraphics[height=5cm]{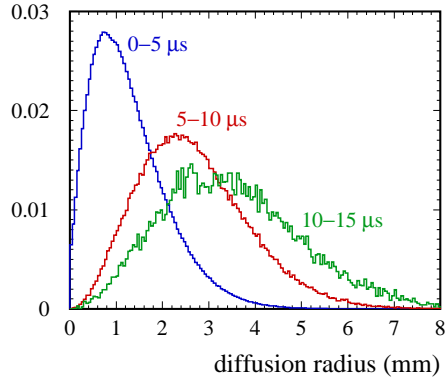}
  \caption{Radial distribution of the $\mu^{-}$~decays for the time
    intervals 0--5, 5--10 and 10--15\un{\mu{}s} at temperature
    $T=300\un{K}$.}
  \label{fig:xdm300}
\end{figure}
The mean diffusion range, which is important for the optimisation of the
pressure and temperature of \chem{H_2} filling the time-projection
chamber, equals about 1\un{mm}. However, long-lived
($t\gtrsim{}10\un{\mu{}s}$) muons travel much farther, which limits the
reachable spatial resolution. Figure~\ref{fig:rdiff} shows the fraction
of the muon decays \textit{outside} the diffusion radius from the point
of $p\mu$~formation, for the target temperature of~30~and~300\un{K}. The
thermal diffusion is significantly reduced at~30\un{K}. This effect is,
however, limited because of the above-mentioned $p\mu$ acceleration in
the spin-flip process. The radial distribution of the muon decays for
several time intervals is plotted in~fig.~\ref{fig:xdm300}.

The calculated values of the mean diffusion radius for a~pure~\chem{H_2}
target at $\phi=0.01$ are summarised in~table~\ref{tab:radius}. The
results are given for $T=30$ and~300\un{K}. The realistic two-Maxwell
distributions of the initial $p\mu$ energies have been used. Also, the
thermalized initial distributions of $p\mu$ atoms with the depleted
$F=1$ state have been employed in order to investigate the thermal part
of the diffusion.
\begin{table}[htb]
  \begin{center}
    \caption{The calculated mean diffusion radius of the $p\mu$ atom in
      pure~\chem{H_2} targets for various initial conditions 
      and~$\varphi=0.01$.}
    \label{tab:radius}
    \begin{ruledtabular}
    \begin{tabular}{cccccc}
      \multicolumn{1}{c}{Temperature}&
      \multicolumn{1}{c}{Initial $\varepsilon$ distribution}&
      \multicolumn{4}{c}{mean~$R_\text{diff}$ [mm] for the time
      interval:} \\
      & & 0--5\un{\mu{}s} & 5--10\un{\mu{}s} & 10--15\un{\mu{}s} 
      & 15--20\un{\mu{}s} \\
      \hline
      30\un{K}  & 0.004\un{eV}(50\%)+1\un{eV}(50\%)  & 
      0.68 & 0.88 & 0.99 & 1.11 \\
      300\un{K} & 0.040\un{eV}(50\%)+5\un{eV}(50\%)  & 
      1.27 & 2.67 & 3.50 & 4.26 \\
      30\un{K}  &  thermal,~$F=0$  & 0.23 & 0.51 & 0.68 & 0.82 \\
      300\un{K} &  thermal,~$F=0$  & 1.11 & 2.59 & 3.44 & 4.09 \\
    \end{tabular}
    \end{ruledtabular}
 \end{center}
\end{table}
\begin{figure} 
  \includegraphics[height=5cm]{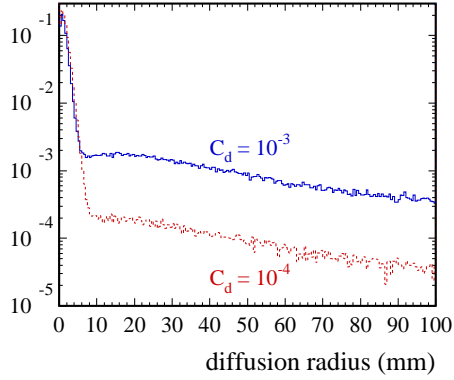}
  \caption{Radial distribution of the $\mu^{-}$~decays in~\chem{H_2}
    with the deuterium concentrations~$C_d=10^{-3}$ and~$10^{-4}$, for
    times $t\leq{}20\un{\mu{}s}$ and $T=300\un{K}$.}
  \label{fig:rdiff_hd}
\end{figure}
\begin{figure} 
  \includegraphics[height=5cm]{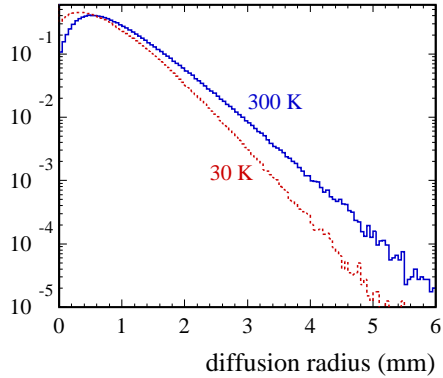}
  \caption{Radial distribution of the $\mu^{-}$~decays from the point of
    $d\mu$ formation in a~pure \chem{D_2}, for times
    $t\leq{}20\un{\mu{}s}$ and $T=30$ and~300\un{K}.}
  \label{fig:rdiff_d2}
\end{figure}
A~real \chem{H_2} target always contains a~certain admixture of
deuterium. Figure~\ref{fig:rdiff_hd} demonstrates that the maximal
muonic atom diffusion radius is greatly increased when the deuterium
concentration of~$10^{-4}$--$10^{-3}$ is present in the
\chem{H_2}~target. The long-range tail in the radial distribution of the
muon decays is due to very energetic ($\approx{}45\un{eV}$) $d\mu$ atoms
formed in the $p\mu$ collisions with deuterons. These $d\mu$ atoms can
travel at large distances owing to the deep Ramsauer-Townsend minimum in
the $d\mu+p$ cross section (see fig.~\ref{fig:xgtran}). Therefore, it is
crucial to reduce the deuterium concentration to a~very low level of
about~$10^{-7}$--$10^{-6}$. The distribution of the $d\mu$ diffusion
radius in a~pure \chem{D_2} gas is shown in~fig.~\ref{fig:rdiff_d2} for
$T=30$ and~300\un{K} and $\varphi=0.01$. The mean value of
$R_\text{diff}$ is smaller (0.80\un{mm} at~300\un{K} and 0.65\un{mm}
at~30\un{K}, for the time interval 0--5\un{\mu{}s}) than in the
pure~\chem{H_2} case since the elastic $d\mu(F=\tfrac{1}{2})+d$ and
$d\mu(F=\tfrac{3}{2})+d$ cross sections are larger than the elastic
$p\mu(F=0)+p$ cross section~\cite{brac89}.  Moreover, the hyperfine
splitting for $d\mu$ is $\Delta{}E^\text{hfs}_{d\mu}=0.0495\un{eV}$. As
a~result, there is practically no spin-flip acceleration of $d\mu$ atoms
at~300\un{K}.  A~relatively weaker acceleration, compared to the
$p\mu+\chem{H_2}$ case, takes place at~30\un{K}.

The Monte Carlo results can be compared with a~simple analytical
estimation. The kinetic theory of gases gives the following mean
diffusion radius~$\overline{R}_\text{diff}$ as a~function of time:
\begin{equation}
  \label{eq:r1}
  \overline{R}_\text{diff}^2 = 6Dt, 
\end{equation}
in which $D$ denotes the diffusion coefficient. It is assumed that the
atom survives until the time~$t$. Using the standard definitions from
the kinetic theory of gases:
\begin{equation}
  \label{eq:def_vl}
  D = \frac{vL}{3} \,, \qquad  L = \frac{1}{\sqrt{2}\sigma N} \,,
\end{equation}
where $v$ is the mean atom velocity, $L$~is the mean free path, $\sigma$
stands for the total cross section and $N$ is the number density of
atoms, one has
\begin{equation}
  \label{eq:r2}
  \overline{R}_\text{diff}^2 = \sqrt{2} \frac{vt}{\sigma N} \,.  
\end{equation}
The factor~$\sqrt{2}$ is valid for a~simple model of the hard sphere
collisions. However, the muonic atom scattering from hydrogenic
molecules is strongly anisotropic. Therefore, we use the following
approximation:
\begin{equation}
  \label{eq:r3}
  \overline{R}_\text{diff} \approx 
  \sqrt{\frac{vt}{\overline{\sigma}_\text{tran}N}} \,,  
\end{equation}
where $\overline{\sigma}_\text{tran}$ is the transport cross
section~(\ref{eq:x_tran}) averaged over the thermal motion of the muonic
atoms and of the target molecules. Taking into account the muon lifetime
$\tau_0=2.2\un{\mu{}s}$, we obtain the following estimation of the mean
diffusion radius:
\begin{equation}
  \label{eq:r4}
  \overline{R}_\text{diff} \approx 
  \sqrt{\frac{v\tau_0}{\overline{\sigma}_\text{tran}N}} \,.  
\end{equation}
For $T=300\un{K}$, $\varphi=0.01$ and a~pure \chem{H_2} target we have
$\overline{\sigma}_\text{tran}=20.8\times{}10^{-20}\un{cm^2}$, which
gives $\overline{R}_\text{diff}\approx{}1.1\un{mm}$. The analogous
estimation for $T=30\un{K}$, using
$\overline{\sigma}_\text{tran}=161\times{}10^{-20}\un{cm^2}$, leads to
$\overline{R}_\text{diff}\approx{}0.23\un{mm}$. These analytical values
are in good agreement with the Monte Carlo results calculated assuming
the thermal initial distribution of $p\mu$ energies and zero population
of the $F=1$ state (see the third column in table~\ref{tab:radius}). In
the real case, the diffusion radius is larger owing to the epithermal
diffusion. Let us note that it is very important to use the molecular
differential cross sections for a~correct Monte Carlo simulation of the
thermal part of the diffusion.  The diffusion radius in~\chem{H_2}
occurs to be about two times smaller than in the case when the
corresponding nuclear cross sections are used.

In conclusion, it has been shown that the optimal conditions for studies
of the $\mu^{-}$~capture on the proton inside the ground-state
$p\mu(F=0)$ atom are achieved at the target density $\phi\sim{}0.01$
when the concentration of the deuterium is depleted to the level
of~$10^{-7}$--$10^{-6}$. The mean diffusion radius of the muonic atoms
at these conditions is on the order of~1\un{mm}. It can be significantly
lowered when the target temperature is decreased from 300\un{K}
to~30\un{K}. This effect is, however, limited as a~fraction of the
$p\mu(F=0)$ atoms is epithermal both due to the initial high-energy
component and to the deexcitation of the $F=1$ states. The simulations
of the muon capture experiments require using the differential cross
sections for the muonic atom scattering from hydrogenic molecules. This
is caused by strong molecular-binding and electron-screening effects at
the collision energies below a~few\un{eV}, where the main stage of the
diffusion process takes place.

\begin{acknowledgments}
  Drs. P.~Kammel, V.~E.~Markushin and C.~Petitjean are gratefully
  acknowledged for stimulating and helpful discussions.
\end{acknowledgments}



\begin{thebibliography}{99}


\bibitem{kamm00}
  P.~Kammel \etal, Nucl.~Phys.~A~\textbf{663-664}, 911c (2000).

\bibitem{kamm01}
  P.~Kammel \etal, Hyperfine Interact.~\textbf{138}, 435 (2001).

\bibitem{kamm02} 
  P.~Kammel \etal, \textit{Proceedings of the International Conference 
  on Exotic Atoms, EXA'2002}, Vienna, November 28--30, 2002, 
  preprint \texttt{nucl-ex/0304019}.

\bibitem{laus05}
  B.~Lauss \etal, \textit{Proceedings of the International Conference on
  Exotic Atoms, EXA'2005}, Vienna, February 21--25, 2005, 
  preprint \texttt{nucl-ex/0601004}.

\bibitem{gorr04}
  T.~Gorringe and H.~W.~Fearing, Rev.~Mod.~Phys.~\textbf{76}, 31 {2004}.

\bibitem{gova00}
  J.~Govaerts and J.-L.~Lucio-Martinez, Nucl.~Phys.~A~\textbf{678}, 
  110 (2000).

\bibitem{adam93}
  A.~Adamczak, Hyperfine Interact.~\textbf{82}, 91 (1993).

\bibitem{adam06}
  A.~Adamczak, accepted for publication in Phys.~Rev.~A, 
  preprint \texttt{physics/0608243}.

\bibitem{brac89}
  L.~Bracci \etal, Muon Catalyzed Fusion~\textbf{4}, 247 (1989).

\bibitem{brac89a}
  L.~Bracci \etal, Phys.~Lett.~A~\textbf{134}, 435 (1989).

\bibitem{brac90}
  L.~Bracci \etal, Phys.~Lett.~A~\textbf{149}, 463 (1990).

\bibitem{chic92}
  C.~Chiccoli \etal, Muon Catalyzed Fusion~\textbf{7}, 87 (1992).

\bibitem{adam96}
  A.~Adamczak, At.~Data Nucl.~Data Tables~\textbf{62}, 255 (1996).

\bibitem{abbo97}
  D.~J.~Abbott \etal, Phys.~Rev.~A~\textbf{55}, 214 (1997).

\bibitem{wert96}
  A.~Werthm\"uller \etal, Hyperfine Interact.~\textbf{103}, 147 (1996).


\end{thebibliography}
\end{document}